\documentclass[aps,pra,reprint,groupedaddress,showpacs,longbibliography]{revtex4-1}

\usepackage[latin1]{inputenc}
\usepackage[T1]{fontenc}
\usepackage{amssymb,amsmath} 
\usepackage{graphicx}
\usepackage[normalem]{ulem}

\newcommand{\mi}{\mathrm{i}}

\newcommand{\e}{\mathrm{e}}
\newcommand{\bs}{\boldsymbol}
\newcommand{\dpsi}{\delta\psi}
\usepackage[colorlinks=true,  urlcolor=blue,  linkcolor=blue,  citecolor=blue]{hyperref}

\begin{document}

\title{The cross-over from Townes solitons to droplets in a 2D Bose mixture}

\author{B. Bakkali-Hassani$^{1}$, C. Maury$^{1}$, S. Stringari$^{2}$, S. Nascimbene$^{1}$, J. Dalibard$^{1}$, J. Beugnon$^{1}$}

\affiliation{$^{1}$Laboratoire Kastler Brossel,  Coll\`ege de France, CNRS, ENS-PSL University, Sorbonne Universit\'e, 11 Place Marcelin Berthelot, 75005 Paris, France}
\affiliation{$^{2}$ INO-CNR BEC Center and Dipartimento di Fisica, Universit\`a degli Studi di Trento, 38123 Povo, Italy  }

\date{\today}

\begin{abstract}
When two Bose-Einstein condensates -- labelled 1 and 2 --  overlap spatially, the equilibrium state of the system depends on the miscibility criterion for the two fluids. Here, we theoretically focus on the non-miscible regime in two spatial dimensions and explore the properties of the localized wave packet formed by the minority component 2 when immersed in an infinite bath formed by component 1. We address the zero-temperature regime and describe the two-fluid system by coupled classical field equations. We show that such a wave packet exists only for an atom number $N_2$ above a threshold value corresponding to the Townes soliton state. We identify the regimes where this localized state can be described by an effective single-field equation up to the droplet case, where component 2 behaves like an incompressible fluid. We study the near-equilibrium dynamics of the coupled fluids, which reveals specific parameter ranges for the existence of localized excitation modes.
\end{abstract}

\maketitle

%%%%%%%%%%%%%%%%%%%%%%%%%%%%%%%%%%%%%%%%%%%%%%%%%%%%%%%%%%%%
%%%%%%%%%%%%%%%%%%%%%%%%%%%%%%%%%%%%%%%%%%%%%%%%%%%%%%%%%%%%
%%%%%%%%%%%%%%%%%%%%%%%%%%%%%%%%%%%%%%%%%%%%%%%%%%%%%%%%%%%%
%%%%%%%%%%%%%%%%%%%%%%%%%%%%%%%%%%%%%%%%%%%%%%%%%%%%%%%%%%%%

\section{Introduction}

Mixtures of quantum fluids display novel phenomenology as compared to the single-component case, through the emergence of collective degrees-of-freedom \cite{2008_pethick,2016_pitaevskii}. In this perspective, ultracold atomic gases have opened a new path for the investigation of such many-body problems, especially thanks to the precise control of interactions \cite{2000:Maddaloni,2009:Fukuhara,2014_ferrier-barbut,2017:Roy,2018:Wu,2019:Desalvo,2019:Huang}. In the case of miscible Bose mixtures with two spin components, the dispersion relations of density and spin linear excitations have been studied experimentally \cite{2020_kim, 2022_cominotti}, as well as the nonlinear excitations known as magnetic solitons \cite{2020_ferrari, 2020_raman}. The superfluid character of the spin degree of freedom was also demonstrated by observing undamped spin-dipole oscillations \cite{2018:Fava} and by moving a magnetic obstacle \cite{2021_kim} in such a mixture. Moreover, it has been shown that a coherent coupling between the two components -- with or without momentum transfer -- can modify the bare dispersion relations in a controlled manner \cite{2022_cominotti}, induce spin-orbit coupling to produce a variety of quantum phases \cite{2022_stringari}, and trigger dynamical instabilities \cite{2011_nicklas}.

Even when each isolated component is stable, an instability can occur in a binary mixture when the interaction between the two components is attractive and set above a certain threshold. Close to this threshold, the balance between the dominant energy contributions can lead to subtle phases. For instance, it was predicted in Ref.\,\cite{2015_petrov} that a mixture of repulsive quantum gases in three dimensions (3D) with finely-tuned mutual attraction may lead to self-bound states stabilized by quantum fluctuations. This novel state of matter, known as a quantum droplet, was realized experimentally in \cite{2017_tarruell, 2018_fattori}, and the link between these 3D droplets and 1D solitons was clarified in \cite{2018_tarruell}. Interestingly, the role of quantum fluctuations is known to be enhanced in low dimensions, resulting in quantum droplet states with properties distinct from the 3D case, both at equilibrium \cite{2016_petrov} and close to equilibrium \cite{2021_sturmer}.

%%%%%%%%%%%% FIGURE 1 %%%%%%%%%%%%%%%%%%%%
\begin{figure}[b]
\centering
\includegraphics[width=\columnwidth]{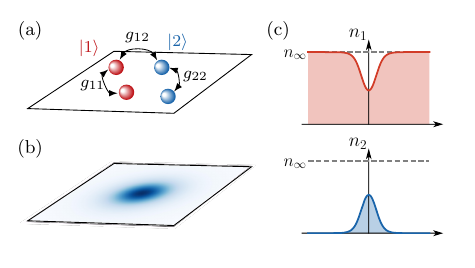}
\caption{Equilibrium state of a 2D minority superfluid component immersed in a bath. (a) Sketch of the interactions in the mixture, with the intra-component couplings $g_{11}$, $g_{22}$ and the inter-component coupling $ g_{12}$. Here, we consider the immiscible regime $g_{12}> \sqrt{g_{11}g_{22}}$. (b) Two-dimensional density profile of the minority component for $m_1=m_2$, $g_{11}=g_{22}$ and $N/N_T = 1.5$, where $N_T\equiv 5.85 / |g_e|$ is the atom number at which the Townes soliton exists and $g_e= g_{22} -  g_{12}^2 / g_{11}$. (c) Density cuts through the center of both components for the same parameters as in (b). At a large distance from the center, the bath density takes the asymptotic value $n_\infty$. }
\label{fig0}
\end{figure}
%%%%%%%%%%%% FIGURE 1 %%%%%%%%%%%%%%%%%%%%

In a different range of parameters, mutually repulsive fluids may experience phase separation, similarly to solutions of Helium 3 in Helium 4 at low temperatures \cite{1978_fetter} or, more prosaically, in a combination of oil and water. This demixing dynamical instability was also characterized using Bose mixtures \cite{1998_timmermans, 1998_hall}. In the regime of strong population imbalance, it was realized early that the dynamics of immiscible mixtures may mimic that of a single closed equation for the minority component \cite{2005_clark, 2013_sartori}. Recently, this mapping was leveraged for the deterministic realization of Townes solitons in a 2D Bose mixture \cite{2021_bakkali}, by making judicious use of the almost coincidence of the various interaction strengths (see also \cite{chen2021observation} for another preparation protocol of the Townes soliton with matter waves and \cite{Kartashov19} for a general review). This approach was also put forward for the realization of other exotic nonlinear excitations, such as Peregrine solitons \cite{2022_romero-ros} or dark-bright soliton trains \cite{2022_romero-ros_b}. 

In this work, we consider a two-component Bose gas and study localized wavepackets of one (minority) component surrounded by a 2D bath of atoms in the other component, see Fig.\,\ref{fig0}. In section \ref{sec:phase_diagram}, we focus on the stationary states of the system. We show that when the atom number in the minority component increases above a threshold value $N_T$, a cross-over transition occurs from a steady-state with a solitonic character (the Townes soliton) to a droplet-like state. Then, in section \ref{sec:excitation_spectrum}, we explore the excitation spectrum of these localized states throughout the crossover. In particular, we show the existence of a given range of atom numbers ($1.45\lesssim N/N_T\lesssim 3.5$) where no localized excitation exist. Finally, we discuss in section \ref{sec:conclusion} some possible extensions of this work.

%%%%%%%%%%%%%%%%%%%%%%%%%%%%%%%%%%%%%%%%%%%%%%%%%%%%%%%%%%%%
%%%%%%%%%%%%%%%%%%%%%%%%%%%%%%%%%%%%%%%%%%%%%%%%%%%%%%%%%%%%
%%%%%%%%%%%%%%%%%%%%%%%%%%%%%%%%%%%%%%%%%%%%%%%%%%%%%%%%%%%%
%%%%%%%%%%%%%%%%%%%%%%%%%%%%%%%%%%%%%%%%%%%%%%%%%%%%%%%%%%%%

\section{Phase diagram}
\label{sec:phase_diagram}

%%%%%%%%%%%% FIGURE 2 %%%%%%%%%%%%%%%%%%%%
\begin{figure*}[t!]
\centering
\includegraphics[width=2\columnwidth,trim={7cm 8cm 7cm 8cm},clip]{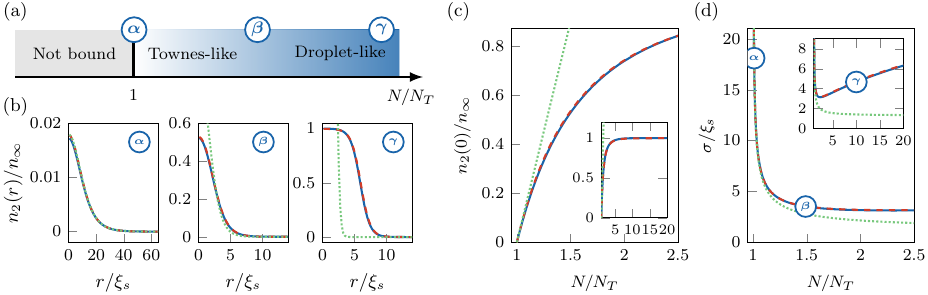}
\caption{(a) Phase diagram of localized states of the minority component. From light to dark blue, one goes from the weak-depletion regime (in which the minority component profile is close to a Townes soliton) to the strong-depletion regime (the central density of the minority component approaches $n_\infty$). (b) Radial density profiles $n_2(r) = |\psi_2(r)|^2$ for selected values of $N/N_T = 1.01$ ($\alpha$), $1.5$ ($\beta$), $10$ ($\gamma$), corresponding to $\mu/\mu_1 \equiv (\mu_2-g_{12} n_\infty)/\mu_1=-7.15\times 10^{-5}, -2.48\times 10^{-3}, -7.62\times 10^{-3}$, respectively. Distances are given in units of the interpenetration length of the immiscible mixture $\xi_s$. (c) Central density as a function of $N / N_T$ (inset: same axes with a different range). (d) RMS size $\sigma$ as a function of $N / N_T$. In (b--d), blue solid lines are obtained from the coupled two-component equations (see Eq.\,\eqref{eq:twocomp_time}) with $g_{12} = 1.01\, g$ with $g\equiv \sqrt{g_{11}g_{22}}$, the green lines correspond to the weak-depletion model of Eq.\,\eqref{eq:rosanov_time} and the red lines refer to the effective single-component model of Eq.\,\eqref{eq:effective_equilibrium}. All profiles are calculated for $m_1=m_2$ and $g_{11}=g_{22}$.}
\label{fig1}
\end{figure*}
%%%%%%%%%%%% FIGURE 2 %%%%%%%%%%%%%%%%%%%%

\subsection{The two-component system}

We consider the ground state of a 2D Bose mixture made of two components, labelled $|1\rangle$ and $|2\rangle$ with mass $m_{1,2}$, and with short-ranged interactions. We use a classical field description of the two components with the two order parameters $\psi_{1,2}(\boldsymbol r,t)$. Both intraspecies and interspecies interactions are assumed to be repulsive, $g_{ij}>0$ where $i,j={1,2}$.  
%with, for simplicity, identical coupling strengths $g_{11}= g_{22}\equiv g > 0$.
%Interspecies interactions are also assumed to be repulsive, $g_{12}>0$. 
In practice, a planar gas is obtained using a strong confinement along $z$. We consider here the quasi-2D situation where the thickness $\ell_z$ is larger than the scattering lengths $a_{ij}$. In such a situation, the classical field approach is valid when $(n_{\rm 3D}a_{ij}^3)^{1/2}\ll 1$, where $n_{\rm 3D}$ is the maximal 3D density of the gas. This hypothesis of negligible beyond-mean-field contributions was well satisfied in the experiment reported in \cite{2021_bakkali}. 

The evolution of $\psi_{1,2}$ is given by the set of coupled nonlinear Schr\"odinger equations (NLSEs)
\begin{equation}
\label{eq:twocomp_time}
\left\{ 
\begin{array}{rl}
\mi \partial_t \psi_1 &= -\frac{1}{2m_1} \nabla^2 \psi_1 +\left( g_{11} |\psi_1|^2 +  g_{12} |\psi_2|^2  \right) \psi_1 \\[5pt]
\mi \partial_t \psi_2 &= -\frac{1}{2m_2} \nabla^2 \psi_2 +\left( g_{22} |\psi_2|^2 +  g_{12} |\psi_1|^2  \right) \psi_2,
\end{array}
\right.
\end{equation}
where $|\psi_i|^2\equiv n_i$ is the 2D atomic density in component $|i\rangle$ (we set $\hbar = 1$). In this work, we look for configurations such that component $|2\rangle$ -- the \textit{minority component} -- contains a finite number of atoms, $N=\int |\psi_2|^2\, \mathrm{d}^2r$, and is localized within the other component $|1\rangle$ -- the \textit{bath} -- which extends to infinity with the asymptotic density $n_\infty$. 

The steady-state of the two-component system is obtained by solving numerically the set of equations \eqref{eq:twocomp_time} for $\psi_i(\boldsymbol r,t) = e^{-\mi\mu_i t}\phi_i(\boldsymbol r)$, where $\mu_{1,2}$ are the chemical potentials of each component. Away from the region where the minority component is localized, the density of the bath brings the energy scale $\mu_1= g_{11} n_\infty$ and the corresponding length scale (healing length) $\xi = 1 / \sqrt{2 m_1 g_{11} n_\infty}$. In pratice, we perform an imaginary time evolution for given $n_\infty$ and $N_2$, from which we determine $\mu_2$. The resulting phase diagram and a few examples of steady-state density profiles are given in Figs.\,\ref{fig1}(a)-\ref{fig1}(b). Before commenting on them, we discuss hereafter various approaches that allow to draw simple physical pictures for this binary mixture.  

%%%%%%%%%%%%%%%%%%%%%%%%%%%%%%%%%%%%%%%%%%%%%%%%%%%%%%%%%%%%
%%%%%%%%%%%%%%%%%%%%%%%%%%%%%%%%%%%%%%%%%%%%%%%%%%%%%%%%%%%%

\subsection{The Townes soliton threshold $N_T$}

When the bath energy scale $\mu_1$ largely exceeds all energy scales governing the minority component dynamics, it is possible to derive a closed equation for the minority component only. This situation is realized when the density of the minority component $n_2$ is everywhere much smaller than the bath density $n_1$, corresponding to a weak depletion of the bath. Under these conditions, atoms in the minority component get dressed by the bath, which induces an additional effective interaction between atoms in state $|2\rangle$ \cite{Bardeen:1967_PhysRev.156.207}. Using Bogoliubov's approach in 2D and for $m_2\gg m_1$, we show in appendix A that this mediated interaction is described by the potential
\begin{equation}
U(\bs{r}) = - \frac{2}{\pi} g_{12}^2 n_\infty K_0 \left( \frac{ r }{ \sqrt{2} \xi } \right),
\label{eq:V2}
\end{equation} 
where $K_0$ is the zeroth-order modified Bessel function of the first kind, with asymptotic behavior $K_0(r) \sim e^{-r} / \sqrt{r}$ when $r \rightarrow +\infty$. This expression is analogous to the Yukawa potential  which arises in a 3D geometry \cite{2008_pethick, 2011_santamore,2018_naidon}. Remarkably, these mediated interactions are always attractive -- whatever the sign of $g_{12}$ -- and their range is given by the bath healing length $\xi$. The same conclusion holds when the masses $m_{1,2}$ are comparable, although the mediated interaction has a more complicated structure in this case (see appendix A and \cite{camacho2018bipolarons,camacho2018landau}).  

When any characteristic length of the minority component is much larger than $\xi$, we can adopt a zero-range description for the mediated interactions. The effective coupling strength, obtained by summing the bare interaction coupling strength $g$ and the mediated one, is then independent of the ratio $m_2/m_1$ and is given by (see Appendix A)
\begin{equation}
g_e = g_{22} - \frac{g_{12}^2} {g_{11}}.
\label{eq:ge}
\end{equation}
In this limit, the minority  component time evolution can thus be approximated -- at least for short times -- by the following single-component NLSE (up to a constant energy contribution)
\begin{equation}
\label{eq:cubic_time}
\mi \partial_t \psi_2 = - \frac{1}{2m_2} \nabla^2 \psi_2 + g_e |\psi_2|^2 \psi_2.
\end{equation}	
We deduce that, in this weak-depletion regime, the existence of stationary localized states for component $|2\rangle$ requires effective attractive interactions, i.e.~$g_e < 0$. This last condition is equivalent to the criterion for the immiscibility $g_{12} > g\equiv \sqrt{g_{11}g_{22}}$ of the binary mixture. The weakly-depleted state is thus a precursor to an actual phase separation situation. 

Eq.\,\eqref{eq:cubic_time} is known to host the so-called Townes soliton \cite{1964_chiao, 1964_talanov,Kartashov19}. Mathematically, this soliton is the unique radially symmetric, real and node-less solution of the stationary version of Eq.\,\eqref{eq:cubic_time}. It exists only when the atom number in the minority component $N $ equals the critical value $N_T \equiv G_T / |g_e|$ with $G_T \simeq 5.85$. When this condition is satisfied, the Townes soliton can be formed with any size, a direct consequence of the scale invariance of Eq.\,\eqref{eq:cubic_time} which does not feature any explicit length scale \cite{2019_saint-jalm}. Formally, the soliton size is set by an effective chemical potential $\mu < 0$ associated to the stationary solution $\psi_2(\boldsymbol r,t)=e^{-\mi\mu t}\phi_2(\boldsymbol r)$ of Eq.\,\eqref{eq:cubic_time} with $\mu = \mu_2 - g_{12} n_\infty$. The upper limit $\mu=0$ is obtained in the case of small depletion, i.e. $n_2\ll n_1\approx n_\infty$ everywhere. The validity of the single-component description of Eq.\,\eqref{eq:cubic_time} was demonstrated experimentally in Ref.\,\cite{2021_bakkali} in this limit.

%%%%%%%%%%%%%%%%%%%%%%%%%%%%%%%%%%%%%%%%%%%%%%%%%%%%%%%%%%%%
%%%%%%%%%%%%%%%%%%%%%%%%%%%%%%%%%%%%%%%%%%%%%%%%%%%%%%%%%%%%

\subsection{The weak-depletion limit}

The scale invariance of Eq.\,\eqref{eq:cubic_time} results from the assumption that the size $\ell$ of the minority component is large compared to the bath healing length $\xi$, which provides the range of the mediated interaction. The first-order correction to this assumption adds a weak nonlocal nonlinearity to Eq.\,\eqref{eq:cubic_time}, which explicitly breaks scale invariance and leads to the modified NLSE (see Appendix\,B)
\begin{equation}
\label{eq:rosanov_time}
\mi \partial_t \psi_2 = - \frac{1}{2m_2} \nabla^2 \psi_2 + g_e |\psi_2|^2 \psi_2 + \beta \left( \nabla^2 |\psi_2|^2 \right) \psi_2,
\end{equation}
with $\beta = -  (g_{12} / g_{11})^2 / (4 m_1 n_\infty)$.  This correction remains small in front of the two dominant terms as long as $m_2|\beta| n_2 \ll 1$ and $\ell \gg \xi_s$ where $\xi_s = (g_{12}/g_{11}) / \sqrt{2 m_1 |g_e| n_\infty}$ represents the interpenetration length -- or ``spin'' healing length -- of the immiscible mixture.

Eq.\,\eqref{eq:rosanov_time} was studied extensively in Ref.\,\cite{2002_rosanov}. The steady state associated with this equation results from the balance between three energetic contributions: (i) the kinetic energy per particle $1/m_2\ell^2$, (ii) the main part of the interaction energy $- N / (N_T m_2\ell^2)$ that balances kinetic energy irrespective of $\ell$ for $N = N_T$, and (iii) the correction $- \beta N / \ell^4$ originating from the extra term in Eq.\,\eqref{eq:rosanov_time} in comparison with Eq.\,\eqref{eq:cubic_time}. For $0 < \epsilon \equiv (N - N_T) / N_T \ll 1$, this balance is achieved for $\ell^2\sim m_2|\beta| N / \epsilon$ and $|\mu| \sim 1 / m_2\ell^2$. 
The extension of these states thus becomes very large when $\epsilon\to 0$, i.e. $N\to N_T^+$, and their density profile  approaches the Townes soliton solution of Eq.\eqref{eq:cubic_time}.
More quantitatively, it is shown in Ref.\,\cite{2002_rosanov} that
\begin{equation}
\label{eq:N_perturbation}
N \approx N_T \left( 1 + 5.43 \, m_2\xi_s^2 |\mu| \right).
\end{equation}

Localized states of Eq.\,\eqref{eq:rosanov_time} exist for any $N > N_T$, by contrast to Eq.\,\eqref{eq:cubic_time} that requires $N=N_T$. When $N$ is close to $N_T$, we recover with this simple approach the phase diagram of Fig.\,\ref{fig1}(a) as well as the density profiles calculated numerically using Eq.\,\eqref{eq:twocomp_time}  (Fig.\,\ref{fig1}, case $\alpha$). For larger $N/N_T$ (Fig.\,\ref{fig1}, cases $\beta$ and $\gamma$), the validity condition $\ell \gg \xi_s$ breaks down, the solution of  Eq.\,\eqref{eq:rosanov_time} is notably different from the result derived from Eq.\eqref{eq:twocomp_time}, and is therefore not relevant for our problem.

%%%%%%%%%%%%%%%%%%%%%%%%%%%%%%%%%%%%%%%%%%%%%%%%%%%%%%%%%%%%
%%%%%%%%%%%%%%%%%%%%%%%%%%%%%%%%%%%%%%%%%%%%%%%%%%%%%%%%%%%%

\subsection{The strong-depletion limit}

When the atom number of the minority component $N$ becomes much larger than $N_T$, the central density  of this component grows to $\bar n_2 = n_\infty\sqrt{g_{11}/g_{22}}$ and the bath is locally fully depleted, see Fig.\,\ref{fig1}(b), case $\gamma$, and Fig.\,\ref{fig1}(c). In this phase-separated regime, the pressures  $g_{ii}n_i^2/2$ in the two components are equal \cite{2008_pethick,2016_pitaevskii}, and component $|2\rangle$ fills approximately  uniformly a disk of radius $R$ such that $N \simeq \pi R^2 \bar n_2$. It thus forms an effective droplet similar to an incompressible fluid of fixed density, although this density is not intrinsic but imposed by the surrounding medium. In the general case, no simple approach is available to describe this regime and one should solve Eqs.\,\eqref{eq:twocomp_time}. Nevertheless, we show in the next paragraph that, close to the SU(2)-symmetry point, the system's equilibrium state can still be described by a single-component equation.

%%%%%%%%%%%%%%%%%%%%%%%%%%%%%%%%%%%%%%%%%%%%%%%%%%%%%%%%%%%%
%%%%%%%%%%%%%%%%%%%%%%%%%%%%%%%%%%%%%%%%%%%%%%%%%%%%%%%%%%%%

\subsection{The vicinity of SU(2) symmetry}

We assume in this paragraph equal masses $m_1=m_2\equiv m$. The interactions are said to be SU(2) symmetric when all interaction parameters $g_{ij}$ are equal. Close to this point, i.e. when $g_{12} \rightarrow g^+$, the stationary state of the mixture in the $N> N_T$ case can be determined by solving the single-component effective equation \cite{2021_bakkali}
\begin{equation}
\label{eq:effective_equilibrium}
\mu \psi_2 = - \frac{1}{2m} \nabla^2 \psi_2 + g_e |\psi_2|^2 \psi_2 + \frac{1}{2m} \frac{ \nabla^2 \sqrt{ n_\infty - |\psi_2|^2 } }{  \sqrt{ n_\infty - |\psi_2|^2 } } \psi_2.
\end{equation}
The data in Figs.\,\ref{fig1}(b)-\ref{fig1}(c) have been calculated for this regime of nearby coupling constants ($g_{12} = 1.01 g$). They show that the predictions derived from Eq.\,\eqref{eq:effective_equilibrium} accurately describe the equilibrium profiles, from the weak to the full depletion regime.

The comparison of the validity ranges for Eq.\,\eqref{eq:effective_equilibrium} and for the stationary version of Eq.\,\eqref{eq:rosanov_time} is instructive: both equations coincide when $|\psi_2|^2\ll n_\infty$ and $g_{12}\approx g$ (Fig.\,\ref{fig1}(b), case $\alpha$). Beyond this common validity domain, Eq.\,\eqref{eq:rosanov_time} allows one to address the case where $g_{12}$ differs notably from $g$, whereas Eq.\,\eqref{eq:effective_equilibrium} is valid for arbitrary depletions, hence arbitrary values of $N/N_T$. Finally we note that one should refrain from using in the general case a time-dependent version of Eq.\,\eqref{eq:effective_equilibrium}, which would be obtained by replacing the left-hand-side $\mu \psi_2$ by $\mi\partial_t\psi_2$. Indeed, in the strong-depletion regime, there is no hierarchy between the time scales for the minority component and for the bath. Therefore, it is not possible to eliminate the bath dynamics and obtain a time-dependent equation for $\psi_2$ involving only a first-order time derivative.

Another remarkable situation which occurs in this SU(2) limit is the  1D dark-bright soliton, introduced by Manakov \cite{Manakov74} and transposed by Busch and Anglin for a binary mixture of Bose gases \cite{Busch01}. There, the majority component wavefunction exhibits a phase jump, akin to a dark soliton around which the minority component accumulates. The first observations of such solitons with cold atoms were reported  in Ref.\,\cite{Anderson01,Becker08}. In a 2D configuration, the equivalent situation would correspond to a vortex texture in the majority component with its core filled by the minority one. This ``vortex-bright" soliton \cite{Anderson01b,Law10} is notably different from the stationary states explored in this article where each component exhibits a uniform phase.

%%%%%%%%%%%%%%%%%%%%%%%%%%%%%%%%%%%%%%%%%%%%%%%%%%%%%%%%%%%%
%%%%%%%%%%%%%%%%%%%%%%%%%%%%%%%%%%%%%%%%%%%%%%%%%%%%%%%%%%%%
%%%%%%%%%%%%%%%%%%%%%%%%%%%%%%%%%%%%%%%%%%%%%%%%%%%%%%%%%%%%
%%%%%%%%%%%%%%%%%%%%%%%%%%%%%%%%%%%%%%%%%%%%%%%%%%%%%%%%%%%%
%%%%%%%%%%%%%%%%%%%%%%%%%%%%%%%%%%%%%%%%%%%%%%%%%%%%%%%%%%%%
%%%%%%%%%%%%%%%%%%%%%%%%%%%%%%%%%%%%%%%%%%%%%%%%%%%%%%%%%%%%
%%%%%%%%%%%%%%%%%%%%%%%%%%%%%%%%%%%%%%%%%%%%%%%%%%%%%%%%%%%%
%%%%%%%%%%%%%%%%%%%%%%%%%%%%%%%%%%%%%%%%%%%%%%%%%%%%%%%%%%%%

\section{Excitation spectrum}
\label{sec:excitation_spectrum}

We now turn to the dynamics of the localized component, restricting for simplicity to close-to-equilibrium  phenomena. This problem goes by essence beyond the Townes soliton physics. Indeed it is known that the Townes soliton associated to Eq.\,\eqref{eq:cubic_time} does not possess any localized mode with non-zero frequency \footnote{There always exists localized modes with zero frequency that are generated by the symmetries of Eqs.\,\eqref{eq:twocomp_time} and which we ignore in the following \cite{1991_malkin}}. More dramatically, some arbitrarily small deformations, such as the multiplication by a phase factor $e^{\mi \alpha r^2}$ with $\alpha\to 0$, may lead to a collapse of the soliton. 

%%%%%%%%%%%% FIGURE 3 %%%%%%%%%%%%%%%%%%%%
\begin{figure*}[t!]
\centering
\includegraphics[width=1.8\columnwidth]{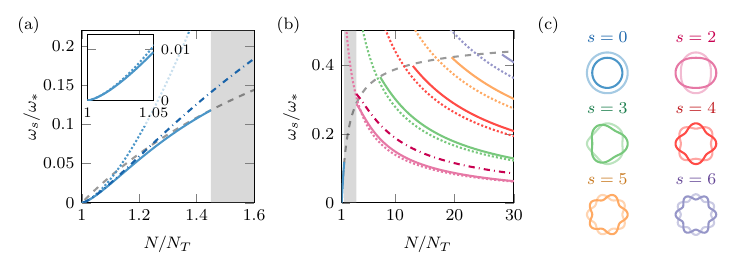}
\caption{Frequencies of localized modes with azimuthal number $s$, for $m_1=m_2$ and for nearby interaction parameters, here $g_{12} = 1.01 g$. The solid lines in (a,b) show the predictions of the Bogoliubov approach for the two coupled equations \eqref{eq:twocomp_time}. (a) Breathing mode $s=0$. The dotted blue line gives the perturbative limit of Eq.\,\eqref{eq:breathing_rosanov}. The inset (same axis) shows that this limit is approached as $N \rightarrow N_T^+$. (b) Surface modes $s\geq 2$. The dotted lines show the hydrodynamic prediction of Eq.\,\eqref{eq:surface_hydro}, see (c) for the color code. In (a) (resp.(b)) the dash-dotted line shows the sum-rule prediction for $s=0$ (resp. $s=2$), while the dashed grey line indicates the limit for localized excitations $\omega_s=|\mu|$. There are no localized modes in the grey-shaded regions of (a) and (b).}
\label{fig2}
\end{figure*}
%%%%%%%%%%%% FIGURE 3 %%%%%%%%%%%%%%%%%%%%

For the two-component system of interest here, a natural approach to determine its excitation spectrum is provided by the Bogoliubov method applied to the coupled equations \eqref{eq:twocomp_time}. This procedure is outlined in Appendix\,C and the results are indicated with full lines in Figs.\,\ref{fig2}(a)-\ref{fig2}(b). Note that we focus here on {localized modes}, i.e. Bogoliubov modal functions $u_i(\boldsymbol r),v_i(\boldsymbol r)$ that decay exponentially to zero for $r\to\infty$. We show in Appendix\,C that this constraint corresponds to a mode frequency $\omega$ smaller than the continuum set by $|\mu|$, where $\mu$ is the effective chemical potential introduced above. We now discuss the results for the various relevant regimes. We consider simple approaches for the limiting cases of weak and strong depletion of the bath and also investigate the intermediate regime of self-evaporation, corresponding to the absence of localized excitations. For simplicity, we restrict in this section to the case of equal masses $m_1=m_2\equiv m$ and equal intracoupling constants $g_{11}=g_{22}=g$.

%%%%%%%%%%%%%%%%%%%%%%%%%%%%%%%%%%%%%%%%%%%%%%%%%%%%%%%%%%%%
%%%%%%%%%%%%%%%%%%%%%%%%%%%%%%%%%%%%%%%%%%%%%%%%%%%%%%%%%%%%

\subsection{Weak-depletion regime and breathing mode}

In the weak-depletion regime, we expect from the results of the previous section that the dynamics of the minority component is well captured by Eq.\,\eqref{eq:rosanov_time}, which takes into account the first-order corrections originating from the two-component nature of our system. Quite remarkably, the instability inherent to Eq.\,\eqref{eq:cubic_time} does not occur for Eq.\,\eqref{eq:rosanov_time}. Indeed, it was shown in Ref.\,\cite{2002_rosanov} that the stationary solutions of Eq.\,\eqref{eq:rosanov_time} are dynamically stable. Moreover, these solutions can sustain a breathing mode, i.e. an oscillation of the system's overall size, for any $N / N_T > 1$. This breathing mode is the only localized mode present for Eq.\,\eqref{eq:rosanov_time}. It may slowly decay (in a non-exponential way) because of nonlinear couplings with excitations in the continuum \cite{1998_afanasjev}. Its frequency $\omega_0$ can be obtained through perturbation theory for small $|\mu|$'s \cite{2002_rosanov}
\begin{equation}
\label{eq:breathing_rosanov}
\omega_0  = 0.95 \, \omega_* \left( N / N_T - 1 \right)^{3/2},
\end{equation}
with $\omega_* = (g / g_{12})^2 |g_e| n_\infty$. This prediction is shown in Fig.\,\ref{fig2}(a) (dotted line). In the limit of small depletion (typically up to $N/N_T<1.05$), it matches well the results of the Bogoliubov analysis. For larger depletions, Eq.\,\eqref{eq:breathing_rosanov} fails reproducing our results. This was expected since the stationary density profile predicted by Eq.\,\eqref{eq:rosanov_time} differs significantly from the exact one in this case.

An upper bound for the breathing mode frequency can be obtained via sum rules \cite{2016_pitaevskii}. This general approach is known to provide acurate estimates of the excitation spectrum for superfluid binary mixtures \cite{2014_ferrier-barbut}. As detailed in Appendix\,D, a relevant sum rule is obtained by looking at the static response of the minority component to a loose harmonic potential (energy weighted sum rule). We also show the result obtained by this method in Fig.\,\ref{fig2}(a).

%%%%%%%%%%%%%%%%%%%%%%%%%%%%%%%%%%%%%%%%%%%%%%%%%%%%%%%%%%%%
%%%%%%%%%%%%%%%%%%%%%%%%%%%%%%%%%%%%%%%%%%%%%%%%%%%%%%%%%%%%

\subsection{Strong-depletion regime and surface modes}

For sufficiently large atom numbers, i.e. in the droplet regime corresponding to an  almost full depletion of the bath, the two-component Bogoliubov analysis shows that there exist localized modes different from the breathing mode (solid lines in Fig.\,\ref{fig2}(b)). More precisely, a quadrupole mode (azimuthal number $s = 2$) detaches from the continuum for $N / N_T \gtrsim 3.5$, see Fig.\,\ref{fig2}(b), and other modes with larger values of $s$ emerge for even larger values of $N/N_T$. We find that the localization for a mode of azimuthal number $s \in \mathbb{N}$ (see Fig.\,\ref{fig2}(c)) approximately occurs when the perimeter of the domain equals $s$ times the spin healing length $\xi_s$, which suggests an interpretation in terms of surface deformations, also called ripplons. 

Such ripplons are well known from 3D incompressible hydrodynamics \cite{1981_landau}. For a two-dimensional system, surface excitations of an incompressible circular bubble of radius $R$ oscillate with an angular frequency $\omega_s$ given by (see e.g.~Ref.\,\cite{1998_nesterov})
\begin{equation}
\label{eq:surface_hydro}
\omega_s = \sqrt{ \frac{\mathcal{T}}{(m_1 + m_2) n_\infty R^3 } s ( s - 1 ) ( s + 1 ) }.
\end{equation}
Eq.\,\eqref{eq:surface_hydro} features a linear tension coefficient $\mathcal{T}$, which has a simple expression in the limit of nearby interaction parameters and equal masses $m_1=m_2\equiv m$ (see e.g.~Refs.\,\cite{1998_ao, 2002_barankov})
\begin{equation}
\label{eq:linear_tension}
\mathcal{T} \simeq \frac{1}{2\sqrt{m}} \sqrt{|g_e| }\; n_\infty^{3/2}.
\end{equation}
In the short-wavelength limit, one retrieves the dispersion relation $\propto k^{3/2}$ with wave number $k = s / R$ expected for a linear (not-curved) interface subject to capillary waves \cite{2002_barankov}. In Fig.\,\ref{fig2}(b), we show that the surface mode frequencies estimated using Eq.\,\eqref{eq:surface_hydro} asymptotically approach the frequencies obtained for large $N / N_T$ from the two-component Bogoliubov approach. 

%%%%%%%%%%%%%%%%%%%%%%%%%%%%%%%%%%%%%%%%%%%%%%%%%%%%%%%%%%%%
%%%%%%%%%%%%%%%%%%%%%%%%%%%%%%%%%%%%%%%%%%%%%%%%%%%%%%%%%%%%

\subsection{The intermediate regime: self-evaporation}

For our choice of nearby interaction parameters, we found that the steady-states comprised in the range $1.45 \lesssim N / N_T \lesssim 3.5$ do not possess any localized excitation mode. Therefore, in this  regime, any perturbation from equilibrium leads to the emission of mass to infinity, a dissipation mechanism known as self-evaporation. This situation is reminiscent of the spectrum of quantum droplets stabilized by beyond mean-field (BMF) effects \cite{2015_petrov, 2020_ferioli, 2021_fort}, as well as of giant resonances observed in nuclear physics \cite{1979_stringari}. 

As discussed in Ref.\,\cite{2020_ferioli}, self-evaporation is not the dominant dissipation mechanism for BMF droplets, because of the prevalence of three-body losses in these large density systems. In contrast, for the two-component mixture considered here, the density of the localized component and thus the three-body loss rate can be tuned through the bath density. For a low-enough density, self-evaporation can then play a relevant role in the damping of the excitations of the system. It could be evaluated either solving explicitly the time-dependent nonlinear Schr\"odinger equations \eqref{eq:twocomp_time} or the extended RPA-Bogoliubov approach accounting for the coupling to the continuum (see e.g.~Ref.\,\cite{1976:Giai}).

%%%%%%%%%%%%%%%%%%%%%%%%%%%%%%%%%%%%%%%%%%%%%%%%%%%%%%%%%%%%
%%%%%%%%%%%%%%%%%%%%%%%%%%%%%%%%%%%%%%%%%%%%%%%%%%%%%%%%%%%%
%%%%%%%%%%%%%%%%%%%%%%%%%%%%%%%%%%%%%%%%%%%%%%%%%%%%%%%%%%%%
%%%%%%%%%%%%%%%%%%%%%%%%%%%%%%%%%%%%%%%%%%%%%%%%%%%%%%%%%%%%

\section{Conclusions and perspectives}
\label{sec:conclusion}

We have presented in this paper a mean-field study of the crossover from a solitonic to a droplet-like behavior in a 2D immiscible Bose mixture. We have determined both the steady-state of the system and its dynamics resulting from a small deviation from equilibrium. We have also proposed simple models that have allowed us to interpret the results obtained in the different limiting regimes.

Regarding the weak-depletion regime, we have shown in Eqs.\,\eqref{eq:rosanov_time} and \eqref{eq:breathing_rosanov} that the interaction mediated by the bath leads to a breaking of the scale invariance of the Townes soliton when its finite range is taken into account. The experimental observation of this emergent length scale should provide a way to discriminate between the scenario studied here and beyond-mean-field effects, i.e. quantum fluctuations, that also provide a mechanism for the stabilization of the minority component wave packet.

The study of the excitation spectrum of the system has revealed the existence of an interval for atom number ($1.45 \lesssim N / N_T \lesssim 3.5$) over which no localized mode exists. This opens the possibility to study the intriguing phenomenon of self-evaporation, in a low-density regime for which other decay mechanisms may be minimized. 

Other future directions of study include the setting in motion of the localized component \cite{2011_saito, 2022_ma}, its link to superfluidity, and the emergence of a roton mode due to a capillary instability \cite{2011_saito_b}. Further clarification on the role of quantum fluctuations close to the miscibility threshold and its influence on the soliton formation may also provide additional interest, as recently discussed for immiscible mixtures in other configurations \cite{2021_naidon}.

\begin{acknowledgments}
This work is supported by ERC TORYD, European Union's Horizon 2020 Programme (QuantERA NAQUAS project) and the ANR-18-CE30-0010 grant. We acknowledge fruitful discussions with D. Petrov, G. Chauveau, F. Rabec and G. Brochier. We thank R. Kaiser for pointing out a related work in the context of photonics \cite{Da22}.
\end{acknowledgments}

%%%%%%%%%%%%%%%%%%%%%%%%%%%%%%%%%%%%%%%%%%%%%%%%%%%%%%%%%%%%
%%%%%%%%%%%%%%%%%%%%%%%%%%%%%%%%%%%%%%%%%%%%%%%%%%%%%%%%%%%%
%%%%%%%%%%%%%%%%%%%%%%%%%%%%%%%%%%%%%%%%%%%%%%%%%%%%%%%%%%%%
%%%%%%%%%%%%%%%%%%%%%%%%%%%%%%%%%%%%%%%%%%%%%%%%%%%%%%%%%%%%

\section*{Appendices}

\subsection{The two-body problem inside a BEC}
\label{subsec:appendix_2_body}

We consider here two impurity atoms immersed in a 3D or 2D uniform BEC (see Fig.\ref{figs1}), and we derive at the lowest relevant order the expression of their effective interaction due to their coupling to the bath. The study of the coupling between one impurity and a bath formed by a BEC, the so-called Bose polaron, is a well-documented problem, see \cite{2015:Grusdt} and refs.\,in. The interaction between two Bose polarons was recently addressed in \cite{camacho2018bipolarons,camacho2018landau,2018_naidon}. We will thus keep our treatment quite brief, and focus on the specificity of the problem addressed in this article.

\paragraph{Yukawa potential for fixed impurities.} We first consider two impurities of infinite mass located in $\bs{R}_a$ and $\bs{R}_b$. They interact with the $N_1$ atoms of the bath by the contact interaction $V=g_{12} \sum_{i=1}^{N_1}\sum_{j=a,b} \delta(\bs{r}_i-\bs{R}_j)$. Using the second-quantized formalism for the bath variables, this interaction reads $V= V_a+V_b$ with
\begin{equation}
V_j = \frac{g_{12}}{\Omega}\sum_{\bs k,\bs k'} a_{\bs k'}^\dagger a_{\bs k}\, {\rm e}^{\mi (\bs k-\bs k')\cdot\bs R_j}.
\label{eq:hat_V_second_q}
\end{equation}
Here $a_{\bs k}$ annihilates a particle of the bath with momentum $\bs k$ and $\Omega$ denotes the volume (resp. area) of the bath in the 3D (resp. 2D) case. 

We assume that the bath is prepared in the $T=0$, fully condensed state of density $n_\infty=N_1/\Omega$, denoted hereafter $|\Phi_0\rangle$ with energy $E_0$, and that its excitations can be described by the Bogoliubov approach. More precisely, we introduce for each momentum $\bs k\neq 0$ the Bogoliubov operators $b_{\bs k}$, $b_{\bs k}^\dagger$ which diagonalize the bath Hamiltonian such that 
\begin{equation}
a_{\bs k}=u_k b_{\bs k}+v_k b^\dagger_{-\bs k},
\end{equation}
where $u_k$, $v_k$ are given by \cite{2016_pitaevskii,2008_pethick}
\begin{equation}
u_k, v_k = \pm \left( \frac{k^2 + 2 m_1 g_{11} }{2k \sqrt{ k^2 + 4 m_1 g_{11} } } \pm \frac{1}{2} \right)^{1/2}
\end{equation}
(as in the main text, we set $\hbar=1$).
%$a_{\bs k}=u_k b_{\bs k}+v_k b^\dagger_{-\bs k}$, $u_k=\cosh \alpha_k$, $v_k=\sinh\alpha_k$, $\tanh (2\alpha_k)=-g_{11}n_\infty/(g_{11}n_\infty+\epsilon_k)$ and $\epsilon_k=k^2/2m_1$ (as in the main text, we set $\hbar=1$). 
We can then rewrite the operators $V_j$ introduced above as $V_j\approx g_{12}n_\infty+ V_j'$ with
\begin{equation}
V'_j=\frac{g_{12}\sqrt{n_\infty}}{\sqrt{\Omega}}\sum_{\bs k\neq 0}(u_k+v_k)\left(b_{\bs k}^\dagger+b_{-\bs k}\right){\rm e}^{-{\rm i}\bs k\cdot \bs R_j},
%\abel{eq:}
\end{equation}
which is the Fr\"{o}hlich Hamiltonian in a BEC \cite{2015:Grusdt}.

%%%%%%%%%%%% FIGURE appendix %%%%%%%%%%%%%%%%%%%%
\renewcommand{\thefigure}{A\arabic{figure}} 
\setcounter{figure}{0}   % reset the figure counter at 0.
\begin{figure}[t!]
\includegraphics[width=0.6\columnwidth]{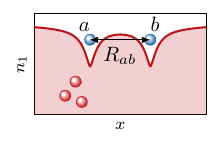}
\caption{Model studied in Appendix A. Two impurities located inside a bath formed by a Bose-Einstein condensate interact by a potential $U(r)$ created by the emission and the absorption of virtual quasi-particles in the bath.}
\label{figs1}
\end{figure}
%%%%%%%%%%%% FIGURE appendix %%%%%%%%%%%%%%%%%%%%

We are interested here in the energy shift of the system that depends on the distance $R_{ab}$ between the particles, and that we will interpret as an effective potential energy between the two impurities. Treating $V$ by perturbation theory, the shift of the ground state energy originating from the $V'_j$'s is up to second order in $g_{12}$ (see \cite{camacho2018landau} for a systematic expansion starting from Eq.\,(\ref{eq:hat_V_second_q})):
\begin{equation}
\Delta E = 2g_{12}n_\infty- \sum_{\alpha\neq0} \frac{
|\langle \Phi_{\alpha} | {V}'_a+ V'_b | \Phi_0  \rangle|^2 }{ E_{\alpha}-E_0 }.
\label{eq:delta_E_1}
\end{equation}
The first contribution is simply the mean-field interaction of each impurity with the BEC.
In the second contribution, the sum runs over all excited states $| \Phi_{\alpha} \rangle$ of the bath. Here, only states with a single excitation $\bs  k$ contribute and we find for
the $R_{ab}$ dependent part of $\Delta E$:
\begin{equation}
U(R_{ab})=-\frac{2g_{12}^2 n_\infty}{\Omega}\sum_{\bs k\neq 0} \frac{(u_k+v_k)^2}{\omega_k}\;{\rm e}^{\mi \bs k\cdot (\bs R_{a}-\bs R_b)}.
\label{eq:Uab}
\end{equation}	
where $\omega_k=\left[\epsilon_k\left(\epsilon_k+2g_{11} n_\infty \right) \right]^{1/2}$ stands for the Bogoliubov dispersion relation of the bath. Note that there are also contributions to $\Delta E$ that do not depend on $R_{ab}$ and that correspond to the self-energy of each  polaron \cite{2015:Grusdt}.

We now turn the discrete sum over $\bs k$ into a $D$-dimensional integral (the fact that the $\bs k=0$ contribution is missing in the sum of Eq.\,(\ref{eq:Uab}) does not play any role for our discussion). We find 
\begin{equation}
U(R_{ab})=-\frac{2g_{12}^2n_\infty}{(2\pi)^D} \int \frac{{\rm e}^{\mi \bs k\cdot \left(\bs R_{a}-\bs R_b\right)}}{\epsilon_k+2g_{11}n_\infty}\;\ {\rm d}^D k,
\label{eq:Delta_E_2}
\end{equation}
i.e. the Fourier transform of a Lorentzian function. It is equal to a Yukawa potential in 3D and to the potential given at Eq. \eqref{eq:V2} in the main text in 2D, with in both cases a range of the order of the bath healing length $\xi=1/\sqrt{2m_1 g_{11} n_\infty}$. In practice, a quasi-2D gas is obtained by starting with a 3D gas and adding a strong confinement along the third direction, with a residual thickness $\ell_z$ of the gas. The 2D version of Eq.\,\eqref{eq:Delta_E_2} then holds when the distance $R_{ab}$ is large compared to $\ell_z$.  

\paragraph{Effective interaction for finite-mass impurities.} When the mass of the impurities $m_2$ is comparable to the mass of the bath particles $m_1$, the kinetic energy of the impurities has to be taken into account in the calculation of the energy shift $\Delta E$. Suppose for example that the two impurities are prepared in a state of well-defined momenta $|\bs p_a,\bs p_b\rangle$. For bosonic impurities, a calculation similar to the one given above leads to the second-order energy shift \cite{camacho2018landau}
\begin{equation}
U(\bs p_a,\bs p_b)=-\frac{g_{12}^2n_\infty}{\Omega}\left(\frac{1}{\omega_k+\Delta}+\frac{1}{\omega_k-\Delta} \right)(u_k+v_k)^2
\label{eq:Delta_E_3}
\end{equation} 
where we have set $k=|\bs p_a-\bs p_b|$ and $\Delta=(\bs p_b^2-\bs p_a^2)/2m_2$. Note that as in Eq.\,(\ref{eq:Uab}), we omitted the self-energy terms which do not play any role in the present work. We  recover the Lorentzian momentum dependence of Eqs.\,(\ref{eq:Uab}-\ref{eq:Delta_E_2}) either if we take the limit $m_2\to \infty$ or, for an arbitrary value of $m_2/m_1$, in the case of a zero center-of-mass momentum $\bs p_b=-\bs p_a$ \cite{camacho2018landau}. An equivalent, alternative approach consists in calculating at second order in $g_{12}$ the scattering amplitude for the collision of two impurities with momenta $\bs p_a,\bs p_b$ in the presence of the bath \cite{camacho2018bipolarons}.
		
\paragraph{Born approximation for the mediated interaction.} For low-temperature gases, it is common to replace the "true" interaction
%, here the mediated potential $U(\bs R)$ given in Eq.\,\eqref{eq:Delta_E_2}, 
by a (regularized) contact interaction $g_{\rm med}\,\delta(\bs R)$, where $g_{\rm med}$ is obtained by taking the zero-energy limit of the scattering amplitude. In this limit, the contribution $\Delta$ in Eq.\,(\ref{eq:Delta_E_3}), which is quadratic with respect to momenta $\bs p_{a,b}$, is negligible in front of $\omega_k$ which varies linearly with $k$. We can then use the result (\ref{eq:Delta_E_2}) obtained for fixed impurities and get:
\begin{equation}
g_{\rm med}= \int U(R)\;{\rm d}^D R=-\frac{g_{12}^2}{g_{11}},
\label{eq:g_med}
\end{equation}	
which is the result used in Eq.\,\eqref{eq:ge}. 

In the 3D case or for the quasi-2D situation where $\ell_z \gg a_{ij}$, the validity of the Born approximation used here requires the scattering length $a_{\rm med}$ associated to the mediated interaction to be much smaller than its range (here $\xi$). Away from a scattering resonance and for $g_{12}\sim g_{11}$, the scattering length for $-\frac{g_{12}^2}{g_{11}}\delta(\bs r)$ is comparable to the van der Waals length, which is indeed much smaller than $\xi$ for a weakly interacting gas. 
%Another formulation of the same constraint is that there should be no bound state in the (attractive) potential  $U(R)$. 
%In 2D, the regularization of the contact interaction for the quantum scattering problem is more subtle than in 3D \cite{Adhikari:1986}, so we will rather discuss the validity of Eq.\,\eqref{eq:g_med} for the classical field approach, after Eq.\,\eqref{eq:non_local_U}.

%One should also keep in mind that the replacement of the mediated interaction by a contact potential is valid only if the characteristic length scale of the problem under study, e.g. the size of the Townes soliton, is large compared to the range $\sim \xi$ of the mediated interaction. Along the same line of thought, we notice that the bath density $n_\infty$ does not enter into $g_{\rm med}$, meaning that a given $g_{\rm med}$ can be reached with an arbitrarily small $n_\infty$. This comes from the fact  that both the strength and the range of $U(R)$ depend on $n_\infty$: When the bath density decreases, the strength of $U(R)$ decreases and its range increases, leaving the integral \eqref{eq:g_med} unchanged. However this integral ceases to be relevant when the range becomes larger than the length scale of the considered problem.

%%%%%%%%%%%%%%%%%%%%%%%%%%%%%%%%%%%%%%%%%%%%%%%%%%%%%%%%%%%%
%%%%%%%%%%%%%%%%%%%%%%%%%%%%%%%%%%%%%%%%%%%%%%%%%%%%%%%%%%%%

\subsection{Adiabatic elimination of the bath field}

Here we consider the two coupled  equations \eqref{eq:twocomp_time} for the classical fields $\psi_{1,2}$ and we explain how they can be simplified into Eq.\,\eqref{eq:rosanov_time} involving only the minority component, when the density $n_2=|\psi_2|^2$ is everywhere small compared to the asymptotic bath density $n_\infty$. 

We recall that the stationary solution of the equation for the bath field $\psi_1$ in the absence of the minority component ($\psi_2=0$) reads $\psi_1(\bs r,t)=\sqrt{n_\infty}\;{\rm e}^{-\mi \mu_1 t}$ with $\mu_1=g_{11}n_\infty$. Here, we treat the field $\psi_2$ in Eqs.\,\eqref{eq:twocomp_time} as a perturbation and we write the field $\psi_1$ as
\begin{equation}
\psi_1(\bs r,t)=\left[\sqrt{n_\infty} +\dpsi_1(\bs r,t)  \right]\,{\rm e}^{-\mi \mu_1 t},
\end{equation}
where $\dpsi_1$ is supposed to be a small correction, meaning that the bath is everywhere only weakly depleted ($n_1(\bs r,t) \approx n_\infty$ everywhere). We now detail how to reduce the initial system of equations to a single closed equation for $\psi_2$:\\

-- First, by keeping all the terms in the first equation of system\,\eqref{eq:twocomp_time} up to order 2 in $\dpsi_1$, we are left with:
\begin{multline}
\label{eq:twocomp_time_app}
g_{12} \sqrt{n_\infty} n_2 + g_{11} n_\infty \left( \dpsi_1 + \dpsi_1^* \right) = \mi \partial_t \dpsi_1 \\ 
- g_{12} n_2 \dpsi_1 - g_{11} \sqrt{n_\infty} \left( 2 |\dpsi_1|^2 + \dpsi_1^2 \right) + \frac{1}{2m_1} \nabla^2 \dpsi_1.
\end{multline}

-- Second, we note that the characteristic time scale for the evolution associated with the minority component is expected to be much longer than the intrinsic time scale $\mu_1^{-1}$ of the bath. Therefore we assume in the following that the state of the bath follows adiabatically the slow motion of the minority component, which amounts to fully neglecting the term $\partial_t \dpsi_1$ in Eq.\,\eqref{eq:twocomp_time_app}. This approximation will be justified \emph{a posteriori} at the end of this appendix.

-- Third, we assume that we can perturbatively expand $\dpsi_1/\sqrt{n_\infty}$ in terms of two small parameters. The first one is $n_2 / n_\infty$ and is associated with the weak-depletion hypothesis mentioned above. The second small parameter is $\xi^2 \nabla^2$ and originates from the fact that the spatial variations of the fields $(\dpsi_1, \psi_2)$ occur on a scale much larger than the bath healing length $\xi$. For clarity, we now introduce the following combinations:
\begin{equation}
\left\{ 
\begin{array}{rl}
\mathcal{S}&= (\dpsi_1 + \dpsi_1^*)/\sqrt{n_\infty} \\[5pt]
\mathcal{D}&= (\dpsi_1 - \dpsi_1^*)/\sqrt{n_\infty}.
\end{array}
\right.
\end{equation}
We obtain the first-order contribution $\mathcal{S}^{(1)}$ to $\mathcal{S}$ by keeping only the terms gathered in the first line of Eq.\,\eqref{eq:twocomp_time_app} (except the time derivative that we dropped):
\begin{equation}
\label{eq:expand_S_1}
\mathcal{S}^{(1)} =  - \frac{ g_{12} }{g_{11}  } \frac{n_2}{n_\infty}.
\end{equation}
To determine $\mathcal{D}^{(1)}$, we consider the difference between Eq.\eqref{eq:twocomp_time_app} and its complex conjugate:
\begin{equation}
-\xi^2\nabla^2 \mathcal{D} +\frac{g_{12}}{g_{11}}\frac{n_2}{n_\infty}\mathcal{D} + \mathcal{S} \mathcal{D} = 0.
%\abel{eq:}
\end{equation}
In this equation valid up to order 2, we can replace $\mathcal{S}$ by its first-order approximation \eqref{eq:expand_S_1}, since it is multiplied by $\mathcal{D}$ which is itself at least of order 1. We therefore obtain at the second order in the small parameters
\begin{equation}
-\xi^2 \nabla^2 \mathcal{D}=0 
%\abel{eq:}
\end{equation}
which implies that $\mathcal{D}^{(1)}= 0$ since we consider only localized perturbations. Using the fact that $\dpsi_1^{(1)}/\sqrt{n_\infty} = (\mathcal{S}^{(1)} + \mathcal{D}^{(1)}) / 2 = \mathcal{S}^{(1)} / 2$, we can then extract the second-order contribution to $\mathcal{S}$ from Eq.\,\eqref{eq:twocomp_time_app}:
\begin{equation}
\label{eq:expand_S_2}
\mathcal{S}^{(2)} = -\left(\frac{g_{12}}{2g_{11}}\frac{n_2}{n_\infty} \right)^2 - \frac{g_{12}}{2g_{11}} \,\xi^2\nabla^2\left(\frac{n_2}{n_\infty}\right).
%- \frac{ g_{12} }{g_{11} \sqrt{ n_\infty } } n_2 - \frac{g_{12}^2}{4 g_{11}^2 n_\infty^{3/2} } n_2^2 - \frac{g_{12}}{4 m_1 g_{11}^2 n_\infty^{3/2} } \nabla^2 n_2.
\end{equation}

-- Finally, we inject the previous results into the equation giving the time evolution of $\psi_2$. More precisely, we expand the density field $n_1$ up to second-order:
\begin{align}
n_1 & = \left| \sqrt{n_\infty} + \dpsi_1 \right|^2 \\
\label{eq:expand_n_1}
& \simeq n_\infty +  n_\infty \left(\mathcal{S}^{(1)} + \mathcal{S}^{(2)} \right) + |\dpsi_1^{(1)}|^2.
\end{align}
This leads to Eq.\,\eqref{eq:rosanov_time}, up to the contribution of the constant energy shift $g_{12}n_\infty$. Note that the first term of the right-hand side of Eq.\,\eqref{eq:expand_S_2} that could give rise to a quadratic dependence in $n_2$ (quintic nonlinearity) eventually cancels with the contribution of $|\dpsi_1^{(1)}|^2$ in Eq.\,\eqref{eq:expand_n_1}. 

One may wonder if it is legitimate to fully neglect the time evolution operator $\partial_t$ in Eq.\,\eqref{eq:twocomp_time_app}, while keeping the laplacian operator in the perturbative expansion. This can be justified a posteriori using the dependence of the breathing frequency $\omega_0$ with the small parameter $\epsilon=N/N_T-1$. This frequency varies as $\epsilon^{3/2}$ (see Eq.\,\eqref{eq:breathing_rosanov}), whereas the wave-packet size is $\propto 1/\sqrt{\epsilon}$. The Laplacian term $\nabla^2\dpsi_1\sim \epsilon \dpsi_1$ is thus large compared to $\partial_t \left( \dpsi_1 \, {\rm e}^{\mi \mu_1 t}\right) \sim \epsilon^{3/2}\left(\dpsi_1\,{\rm e}^{\mi \mu_1 t}\right)$ and the procedure outlined here is legitimate close to the Townes threshold. However, one should expect significant corrections due to the non-adiabatic following of the bath variables as soon as $N$ deviates significantly from $N_T$ (see also the discussion after Eq.\,\eqref{eq:effective_equilibrium}). 

%%%%%%%%%%%%%%%%%%%%%%%%%%%%%%%%
%%%%%%%%%%%%%%%%%%%%%%%%%%%%%%%%
%%%%%%%%%%%%%%%%%%%%%%%%%%%%%%%%
%%%%%%%%%%%%%%%%%%%%%%%%%%%%%%%%

\subsection{Excitations of the two-component system}

We consider the two coupled equations \eqref{eq:twocomp_time} that give the evolution of the two classical fields $\psi_{1,2}$, choosing for simplicity $m_1=m_2\equiv m$. We assume that the minority component contains $N>N_T$ atoms, so that there exists a stable localized state for this component. The steady-state of the system is thus characterized by the real radial wave functions $R_{1,2}(r)$. We look for perturbations around this steady-state by setting

\begin{equation}
\left\{ 
\begin{array}{rl}
	\psi_1(\textbf{r}, t) &= \left[ R_1(r) + \alpha_1( \bs{r}, t) + \mi \beta_1( \bs{r}, t) \right] \e^{ -  \mi \mu_1 t } \\[5pt]
	\psi_2(\textbf{r}, t) &= \left[ R_2(r) + \alpha_2( \bs{r}, t) + \mi \beta_2( \bs{r}, t) \right] \e^{ -  \mi \mu_2 t },
\end{array}
\right.
\end{equation}
where the small perturbations $\alpha_1, \beta_1, \alpha_2, \beta_2$ are by construction real functions. 

The  evolution of the $\alpha_j$ and $\beta_j$ is given by the linear system
\begin{equation}
\partial_t
\begin{pmatrix} \alpha_1 \\ \beta_1 \\ \alpha_2 \\ \beta_2 \end{pmatrix}= 
\begin{pmatrix} 
	0  &  L_0^{(1)}  &  0  &  0  \\
	 -L_1^{(1)}  &   0  &  -L_{12}  & 0 \\ 
	 0  &  0  &  0 &  L_0^{(2)}   \\
	 -L_{12}  & 0  &   -L_1^{(2)}  & 0
 \end{pmatrix}
\begin{pmatrix} \alpha_1 \\ \beta_1 \\ \alpha_2 \\ \beta_2 \end{pmatrix},
\end{equation}
with the following differential operators
\begin{align}
	L_0^{(1)} & = -\mu_1 - \frac{1}{2m} \nabla^2 + g_{11} R_1^2 + g_{12} R_2^2 \\
	L_1^{(1)} & = -\mu_1 - \frac{1}{2m} \nabla^2 + 3 g_{11} R_1^2 + g_{12} R_2^2.
\end{align}
The operators $(L_0^{(2)}, L_1^{(2)})$ are deduced from $(L_0^{(1)}, L_1^{(1)})$ by exchanging the indices $1$ and $2$ in these last equations. We also introduced the operator $L_{12}$ coupling the two components
\begin{equation}
	L_{12} = 2 g_{12} R_1 R_2.
\end{equation}

For $r$ large compared to the extension of the localized component, $L_{12}$ vanishes and one can check that a localized excitation of component $|2\rangle$, varying as ${\rm e}^{-\kappa r} / \sqrt{r}$ for large $r$, will have a frequency $\omega=g_{12}n_\infty-\mu_2-\kappa^2/2m=|\mu|-\kappa^2/2m$, with $\mu=\mu_2-g_{12}n_\infty<0$, whereas a delocalized excitation varying as ${\rm e}^{\pm \mi k r} / \sqrt{r}$ will correspond to $\omega=|\mu| + k^2/2m$ . This means that the condition $\omega<|\mu|$ is a necessary condition for the excitation of component $|2\rangle$ to be localized. Numerically, the localized excitations as shown in Fig.\,\ref{fig2} are identified by noticing that their frequency and functional form do not depend on the extension of the calculation grid.

%%%%%%%%%%%%%%%%%%%%%%%%%%%%%%%%%%%%%%%%%%%%%%%%%%%%%%%%%%%%
%%%%%%%%%%%%%%%%%%%%%%%%%%%%%%%%%%%%%%%%%%%%%%%%%%%%%%%%%%%%

\subsection{Sum-rules}

\paragraph{Monopole mode.}
The general formalism of sum rules provides sharp upper bounds for the excitation spectrum of many-body systems \cite{2016_pitaevskii}. We can estimate the frequency of the monopole breathing mode by calculating the ratio $M_1/M_{-1}$ between the energy-weighted and the inverse-energy-weighted sum rules relative to the operator $F_0 \equiv x^2+y^2$ of the minority component. The energy-weighted sum rule is easily calculated using basic commutator rules:
\begin{equation}
\label{eq:energy_weighted_monop}
M_1 = -2 \langle x^2 + y^2\rangle,
\end{equation}
where the average should be taken by integrating the density of the minority component using  the ground state wave function of the mixture. The inverse-energy-weighted sum rule requires the calculation of the static response of the system $\delta \langle x^2 + y^2\rangle$ to a perturbation of the form $m_2\lambda_0 ( x^2 + y^2 )$ (again applied only to the minority component). One then obtains
\begin{equation}
M_{-1} = - \frac{1}{2\lambda_0} \delta \langle x^2 + y^2\rangle.
\end{equation}
In conclusion, a rigorous upper bound to the frequency of the lowest monopole mode is given by
\begin{equation}
\omega^2_0 \le \frac{M_1}{M_{-1}} = 4 \lambda_0 \, \frac{ \langle x^2 + y^2 \rangle }{ \delta \langle x^2 + y^2 \rangle }.
\end{equation}

\paragraph{Surface modes.}
The same approach can be employed to estimate the surface mode frequencies. For example, the quadrupole mode can be usefully described using the excitation operator $F_2 \equiv x^2 - y^2$. In this case, the energy-weighted sum rule is still given by Eq.\,\eqref{eq:energy_weighted_monop} holding for the monopole excitation, while the inverse-energy-weighted moment requires a calculation based on a perturbation of the type $m_2\lambda_2 (x^2 - y^2)$ applied only to the minority component. The result for the quadrupole frequency is then given by
\begin{equation}
\omega^2_2 \le - 4 \lambda_2 \,\frac{\langle x^2 + y^2\rangle  }{ \delta \langle x^2 - y^2 \rangle}.
\end{equation}
It is easy to check that, when applied to a harmonically-trapped 2D single-component BEC with repulsive interactions, the monopole and quadrupole frequencies estimated above coincide exactly with the hydrodynamic values $\omega_0 = 2 \omega_{\rm ho} $ and $\omega_2 = \sqrt{2} \omega_{\rm ho}$, the latter result holding in the Thomas-Fermi limit. In the calculation of the quadrupole static response, one should pay attention to the fact that the addition of the perturbation $m_2\lambda_2 (x^2 - y^2)$  may induce a collapse of the system at large distances, where the potential becomes deeply attractive along $x$ (or $y$, depending on the sign of $\lambda_2$). The simplest way to evaluate the quadrupole response function while avoiding the risk of collapse is to add a perturbation of the form
\begin{equation}
2m\lambda x^2= m\lambda(x^2+y^2) + m\lambda(x^2-y^2),
\end{equation}
with $\lambda$ small and positive. In this way, there is no collapse at large distances and one can simultaneously calculate both the monopole $\delta \langle x^2+y^2\rangle /\lambda$ and   quadrupole $\delta \langle x^2 - y^2\rangle / \lambda$ responses, thereby giving access to both $\omega_0$ and $\omega_2$ with the same simulation. 

%More generally, a sum rule for surface modes of higher azimuthal numbers $s$ can be obtained by applying a static perturbation of the type $\lambda (F_0 + F_s)$ with $F_s = r^2 \cos( s \theta )$ expressed in terms of the polar coordinates $(r, \theta)$. The energy-weighted sum rule writes now
%\begin{equation}
%m_1  = \left\langle - 2 r^2 \left[ \cos^2 ( s \theta ) + \frac{s^2}{4} \sin^2 ( s \theta ) \right] \right\rangle,
%\end{equation}
%while the inverse energy-weighted sum rule still requires the numerical determination of the static response $\delta \langle F_s \rangle$ of the system. One recovers the previous results for $s = 0$ and $s = 2$.

\bibliography{Spin_bubble_spectrum_bib_v3}
\clearpage
\end{document}